\documentclass[a4paper,11pt]{article}
\usepackage{jcappub} 
\usepackage{lineno}

\title{\boldmath Astrophysical environment around a black hole in the braneworld and its optical signatures}

\author[a,1]{M.~F.~Fauzi,\note{Corresponding author.}}
\author[b]{A.~O.~Latief,}
\author[a]{and A.~Sulaksono}
\affiliation[a]{Departemen Fisika, FMIPA, Universitas Indonesia,\\
Depok 16424, Indonesia}
\affiliation[b]{Physics of Magnetism and Photonics Research Division, Faculty of Mathematics and Natural Sciences, Institut Teknologi Bandung,\\
Jl.~Ganesha no.~10, Bandung 40132, Indonesia}

\emailAdd{muhammad.fahmi31@ui.ac.id}
\emailAdd{latief@itb.ac.id}
\emailAdd{anto.sulaksono@sci.ui.ac.id}

\abstract{
We investigate the impact of braneworld theory on the astrophysical environment surrounding a black hole. The black hole is sourced by localized matter from the bulk, which could describe both regular and singular (Schwarzschild) black hole. Employing an Einstein cluster description for the environment, we find that the anisotropic nature of the cluster, coupled with finite brane tension, leads to a weakening of gravity due to the quadratic and nonlocal corrections to the effective four-dimensional field equations. Consequently, this effect prevents horizon formation within the environment. Applying current constraints on the brane tension derived from neutron star observations, we demonstrate that this effect is particularly relevant for sub-stellar mass black holes embedded in compact environments. Furthermore, we investigate the optical signatures of finite brane tension in this scenario, specifically focusing on the black hole shadow and Einstein ring radii. We show that the Einstein ring radius decreases with a smaller brane tension, whereas the black hole shadow radius increases\textemdash somewhat contradicts the weakening gravity effects. Ultimately, these two observables may jointly serve to constrain the value of the brane tension in a very specific astrophysical scenarios.
}

\begin{document}
\maketitle
\flushbottom

\section{Introduction}

The Randall–Sundrum (RS) braneworld paradigm is one of the most significant extensions of general relativity (GR)~\cite{Randall-Sundrum-large-1999,Randall-Sundrum-alternative-1999,Maartens:2010ar}. It postulates that our observable universe is a brane with tension embedded in a warped higher-dimensional bulk spacetime, within which gravity can propagate. In braneworld theory, the generic black hole (BH) solution takes a form similar to the Reissner–Nordström BH, but allows for a negative $1/r^2$ term in the metric coefficient~\cite{Dadhich:2000am}, where the corresponding ``charge" is known as a \emph{tidal charge}. Several works have investigated signatures of the braneworld model, including the proposal that galactic rotation curves can be explained by bulk effects rather than dark matter (DM)~\cite{Mak-Harko-can-2004,Bohmer-Harko-galactic-2007}. A class of braneworld BH solutions that incorporates DM-like effects in the gravitational potential was studied in Ref.~\cite{Heydar-Fard:2007ahl}. A Schwarzschild-like BH solution in the brane vacuum was recently constructed by Nakas and Kanti~\cite{Nakas:2020sey}, and later extended in Ref.~\cite{Neves:2021dqx} to obtain a regular BH solution within the same formalism. Their construction effectively ``forces" a matter content in the five-dimensional bulk to produce a (regular) BH geometry on the four-dimensional brane, as described by the effective field equations~\cite{Shiromizu:1999wj}. In this framework, there is no explicit matter content on the brane itself; instead, the resulting geometry arises from the influence of higher-dimensional \emph{bulk matter}.

On the observational side, numerous studies have explored the imprint of braneworld physics on astrophysical BH observations, including gravitational lensing~\cite{Zhang:2023rsy,Whisker:2004gq,Horvath:2010xq,Bin-Nun:2010exl,Horvath:2012ru,Abdujabbarov:2017pfw}, BH shadows~\cite{Amarilla:2011fx,Eiroa:2017uuq,Hou:2021okc,Banerjee:2019nnj,Neves:2020doc,KumarSahoo:2025leq}, and gravitational-wave observations~\cite{Toshmatov:2016bsb,McWilliams:2009ym}. Other studies on the properties of compact objects in braneworld theory have also been extensively studied. In particular, Ref.~\cite{Lugones:2017jak} showed that finite brane tension allows for arbitrarily massive neutron stars without a maximum mass. Lower bounds on the brane tension $\sigma$ have been obtained as $\sigma \gtrsim 10^{37}\,\text{dyn/cm}^2$ from neutron-star observations~\cite{GonzalezFelipe:2016edf,Murshid:2025ppz}. See also Refs.~\cite{Germani:2001du,Castro:2014xza,Prasetyo:2017hrb,Hladik:2011zz} for further studies on stars and compact objects in the braneworld.

In parallel with foundational studies in gravitation, interest in the astrophysical environment surrounding BHs has grown significantly in recent years. Indeed, astrophysical BHs are expected to be embedded in matter distributions~\cite{Speri:2022upm,Yunes:2011ws,Gondolo:1999ef,Sadeghian:2013laa,Bertone:2004pz}. This scenario was recently discussed by Cardoso \emph{et al.}~\cite{Cardoso:2021wlq} in the context of a BH surrounded by a matter distribution with vanishing radial pressure\textemdash an \emph{Einstein cluster}, describing a system of self-gravitating particles in circular orbits~\cite{Einstein:1939ms}. This matter distribution is commonly interpreted as DM, given that DM is expected to interact predominantly through gravity~\cite{Bertone:2004pz}, and that the DM energy density profile is typically used to construct the energy–momentum tensor of the Einstein cluster~\cite{Cardoso:2021wlq}. The vanishing radial pressure provides several advantages for theoretical analyses: the deformed spacetime geometry near the BH becomes more tractable, often allowing (approximate) analytical solutions depending on the chosen density profile~\cite{Cardoso:2021wlq,Konoplya:2022hbl}. Moreover, it has been shown that the dominant energy condition and causality (via the speed of sound) may be violated if the DM extends close to the BH horizon~\cite{Datta:2023zmd}.

Recent works have actively explored the observational implications of Einstein clusters surrounding BHs, particularly in both gravitational-wave~\cite{Cardoso:2022whc,Destounis:2022obl,Speeney:2024mas,Figueiredo:2023gas,Gliorio:2025cbh} and electromagnetic (optical) channels~\cite{Kouniatalis:2025itj,Xavier:2023exm,Macedo:2024qky,Fauzi:2025yse,Fernandes:2025osu}. While our focus is on optical observables, previous studies suggest that significant deviations arise only when the surrounding environment is sufficiently compact. Although such compact configurations may not correspond to current astrophysical observations, studying this regime provides valuable qualitative insight into how matter environments modify BH observables.

The interplay between environmental effects and braneworld physics in our observable universe\textemdash and their associated observational signatures, especially in strong-field regimes near BHs\textemdash is particularly compelling. Building on the framework of Refs.~\cite{Nakas:2020sey,Neves:2021dqx}, one can incorporate matter on the brane\textemdash here, an Einstein cluster\textemdash into the effective four-dimensional field equations~\cite{Shiromizu:1999wj} and investigate its impact on both the matter distribution and the resulting spacetime geometry. These modifications, in turn, affect particle dynamics and may leave observable imprints in BH observations and other astrophysical systems.

One optical observable of BHs that has become of particular interest in theoretical studies is the BH shadow radius. Results from the Event Horizon Telescope (EHT) on images of the M87*~\cite{EventHorizonTelescope:2019dse,EventHorizonTelescope:2019uob,EventHorizonTelescope:2019jan,EventHorizonTelescope:2019ths,EventHorizonTelescope:2019pgp,EventHorizonTelescope:2019ggy,EventHorizonTelescope:2021bee,EventHorizonTelescope:2021srq,EventHorizonTelescope:2023gtd} and Sagittarius (Sgr) A*~\cite{EventHorizonTelescope:2022wkp,EventHorizonTelescope:2022apq,EventHorizonTelescope:2022wok,EventHorizonTelescope:2022exc,EventHorizonTelescope:2022urf,EventHorizonTelescope:2022xqj,EventHorizonTelescope:2024hpu,EventHorizonTelescope:2024rju} BHs have motivated numerous studies investigating the shadows of BHs in various theories and scenarios, including both the previously discussed astrophysical environments~\cite{Xavier:2023exm,Kouniatalis:2025itj,Fauzi:2025yse,Macedo:2024qky,Fernandes:2025osu} and braneworld theory~\cite{Banerjee:2019nnj,Neves:2020doc,Hou:2021okc,KumarSahoo:2025leq}. It was found that astrophysical environments lead to an increasing BH shadow radius with increasing mass or compactness of the matter distributions~\cite{Xavier:2023exm,Fauzi:2025yse,Macedo:2024qky,Fernandes:2025osu}. In the context of braneworld BHs with tidal charge, the shadow radius behavior depends on the sign of the tidal charge~\cite{Neves:2020doc}. However, the shadow radius alone may not provide sufficient information to determine the underlying geometry: various static and spherically symmetric BH solutions can give rise to the same shadow radius by tuning their parameters~\cite{Vagnozzi:2022moj}. On the other hand, the shape of the shadow of a rotating BH may tell a different story, as several studies indicate that non-spherically symmetric features in some models and theories can produce distinct deformations in the shadow shape~\cite{Eichhorn:2021etc,Eichhorn:2021iwq,Chen:2025jay}. Nonetheless, such observations may need to be supplemented with probes in the weaker-field regime.

Probing the weaker-field regime using the electromagnetic channel requires observables that originate farther away from the BH. Gravitational lensing is therefore a strong candidate. The regime probed by lensing observables is particularly relevant in the case of BH environments, as it spans large distances from the BH. Lensing effects have been widely used to probe the structure of galaxies~\cite{Keeton:1997by,Vegetti:2023mgp}, and even to determine the Hubble constant through time-delay measurements (see Ref.~\cite{Birrer:2022chj} and the references therein). Moreover, the light deflection caused by a BH surrounded by an Einstein cluster in GR has been investigated recently in Refs.~\cite{Kouniatalis:2025itj,Fauzi:2025yse}, which found that the Einstein cluster generally leads to a stronger light deflection angle (see also Ref.~\cite{Boehmer:2007az}). From the braneworld perspective, similar to the BH shadow radius, the effect of BH tidal charge on the lensing observables depends on the sign of the charge~\cite{Whisker:2004gq,Horvath:2010xq,Abdujabbarov:2017pfw}, and some studies were able to obtain upper and lower bounds on the tidal charge using these observables~\cite{Zhang:2023rsy,Bin-Nun:2010exl,Horvath:2012ru}.

In this study, we aim to investigate, for the first time, the properties of the astrophysical environment surrounding a BH from the perspective of braneworld theory and its observational signatures in the optical channel. Discussed in Sec.~\ref{sec. braneworld BH}, the four-dimensional BH geometry is sourced from the \emph{bulk} without tidal charge, following Refs.~\cite{Nakas:2020sey,Neves:2021dqx}, which can describe both singular and regular BHs. In Sec.~\ref{sec. matter on the brane}, we add matter onto the brane\textemdash the Einstein cluster, as studied in Ref.~\cite{Cardoso:2021wlq}\textemdash and examine how it affects the field equations on the brane. We will see in this section that the vanishing radial pressure of the Einstein cluster significantly simplifies the equations. In Sec.~\ref{sec. result braneworld modification}, we show that the anisotropic characteristics of the Einstein cluster have a significant influence within the braneworld formalism, affecting both the transverse pressure and the spacetime geometry arising from the environment, which constitutes the main focus of our study. Furthermore, in Sec.~\ref{sec. observables}, we investigate the optical signatures, particularly the photon spheres, shadow radius, and the Einstein ring angular radius. Finally, we conclude our findings in Sec.~\ref{sec. conclusion}.

Throughout the paper, we use geometrized units with $G = c = 1$, where $G$ and $c$ are the Newtonian gravitational constant and the speed of light, respectively. We also define $\kappa^2 = 8\pi G$. Additionally, we avoid using $\kappa_5^2$\textemdash a quantity related to the gravitational constant in higher dimensions\textemdash as used in, e.g., Refs.~\cite{Nakas:2020sey,Neves:2021dqx}, since it introduces ambiguity into the equations. Instead, we expand every occurrence of $\kappa_5^2$ based on its relation to the four-dimensional gravitational constant $\kappa^2$ and the brane tension $\sigma$, namely $\kappa_5^4 = \kappa^2 / 6\sigma$~\cite{Maartens:2010ar,Nakas:2020sey,Neves:2021dqx}. We also use uppercase Latin indices ($M$ and $N$) for five-dimensional quantities and Greek indices ($\mu$ and $\nu$) for four-dimensional quantities.

\section{Braneworld (regular) black hole}
\label{sec. braneworld BH}

A BH solution in five-dimensional RS spacetime was proposed by Nakas and Kanti~\cite{Nakas:2020sey}, and later followed by Neves~\cite{Neves:2021dqx} who interpreted a more general regular BH solution from a braneworld perspective. The five-dimensional RS-II spacetime is expressed by the line element
\begin{equation}
    ds^2=e^{-2k|y|}(-dt^2+d\vec{x}^2)+dy^2,
\end{equation}
where $k$ is a quantity related to the anti-de Sitter length of the bulk $\ell_{AdS}$ by $k=1/\ell_{AdS}$, $\vec{x}$ represents the usual three-dimensional spatial coordinates, and $y$ is the additional dimension in the bulk. Equivalently, by defining $z=\text{sgn}(y)\left[\exp(k|y|)-1\right]/k$, we obtain
\begin{equation}
    ds^2=\frac{1}{(1+k|z|)^2}(-dt^2+dr^2+r^2d\Omega_2^2+dz^2).
\end{equation}
One can then impose the coordinate transformation~\cite{Nakas:2020sey}
\begin{equation}
    r=\rho\sin\chi,\qquad z = \rho\cos\chi,
\end{equation}
so that the line element becomes
\begin{equation}
    ds^2=\frac{1}{\left(1+k\rho|\cos\chi|\right)^2}\left(-dt^2+d\rho^2+\rho^2d\Omega_3^2\right),
\end{equation}
where $d\Omega_3^2=d\chi^2+\sin^2\chi d\theta^2 + \sin^2\chi\sin^2\theta d\phi^2$ is the unit three-sphere element.

Now, for the regular BH spacetime, we require the metric to take the form~\cite{Nakas:2020sey}
\begin{equation}
    ds^2=\frac{1}{(1+k\rho\cos\chi)^2}\left(-f(\rho)dt^2+\frac{d\rho^2}{f(\rho)}+\rho^2d\Omega_3^2\right),
    \label{eq. line element 5d}
\end{equation}
where~\cite{Neves:2021dqx}
\begin{equation}
    f(\rho)\equiv1-\frac{2\bar{m}(\rho)}{\rho},
\end{equation}
with $\bar{m}(\rho)$ representing the five-dimensional mass profile of the regular BH. Here, we specifically choose the Hayward-type mass profile~\cite{Hayward:2005gi,Cadoni:2022chn},
\begin{equation}
    \bar{m}(\rho)=\frac{M_{BH}}{1+\left(\ell/\rho\right)^3},
    \label{eq. mass hayward 5D}
\end{equation}
where $M_{BH}$ is the Arnowitt–Deser–Misner (ADM) mass of the BH and $\ell$ is the regularization parameter. Note that the regularization parameter $\ell$ is, in general, not related to the anti-de Sitter length of the bulk $\ell_{AdS}$, although one has the freedom to construct such a relation. Employing the five-dimensional field equation,
\begin{equation}
    ^{(5)}G_{MN}=T_{MN}^{(B)},
    \label{eq. EFE 5D}
\end{equation}
each component of the bulk five-dimensional energy-momentum tensor $T_{MN}^{(B)}=\text{diag}(\epsilon_B,p_1,p_2,p_2,p_2)$, which sources the spacetime metric in the $(t,\rho,\theta,\phi,\chi)$ coordinates, can be determined (see Refs.~\cite{Neves:2021dqx,Nakas:2020sey} for details).

To obtain the projected four-dimensional field equation, one can use the well-known Shiromizu–Maeda–Sasaki procedure~\cite{Shiromizu:1999wj}. Here, we impose the four-dimensional line element ansatz in the form
\begin{equation}
    ds^2=\underbrace{-e^\nu}_{g_{tt}} dt^2+ \underbrace{e^\lambda}_{g_{rr}} dr^2 + r^2d\Omega^2_2,
    \label{eq. metric line element 4d}
\end{equation}
where $e^{-\lambda}=1-2m(r)/r$, with $m(r)=\bar{m}(r,0)$. The resulting four-dimensional field equation reads
\begin{equation}
    G_{\mu\nu}=\kappa^2\left(\tau_{\mu\nu}+T^{(e)}_{\mu\nu}+\frac{6}{\sigma}\pi_{\mu\nu}-\frac{\sigma}{2}g_{\mu\nu}\right) - \mathcal{E}_{\mu\nu},
    \label{eq. field eq brane}
\end{equation}
where $T_{\mu\nu}^{(e)}$ is the induced energy-momentum tensor projected from the five-dimensional bulk onto the brane, while $\tau_{\mu\nu}$ and $\pi_{\mu\nu}$ are, respectively, the energy-momentum tensor for matter on the brane and its high-energy quadratic correction, given by
\begin{equation}
    \pi_{\mu\nu}=\frac{\tau\tau_{\mu\nu}}{12} - \frac{\tau_{\mu\alpha}\tau^{\alpha}_\nu}{4} + \frac{g_{\mu\nu}}{24}\left(3\tau_{\alpha\beta}\tau^{\alpha\beta} - \tau^2\right).
\end{equation}
The constant $\sigma$ is the brane tension, also interpreted as the vacuum energy, and is related to $k$ by $\sigma=6k^2/\kappa^2=6/(\ell_{AdS}^2\kappa^2)$. The projection of the five-dimensional bulk energy-momentum tensor $T^{(B)}_{MN}$ onto the brane is given by
\begin{equation}
T^{(e)}_{\mu\nu}=\frac{2}{3}\sqrt{\frac{6}{\sigma \kappa^2}}
\left[g^M_\mu g^N_\nu T_{MN}^{(B)} \left.+\left(T_{MN}^{(B)}n^Mn^N-\frac{T^{(B)}}{4}\right)g_{\mu\nu}\right]\right|_{y\to0},
\end{equation}
where $n^M$ is a spacelike unit normal to the brane. Lastly, $\mathcal{E}_{\mu\nu}$ is the non-local correction arising from the projection of the bulk Weyl tensor onto the brane,
\begin{equation}
    \mathcal{E_{\mu\nu}}=C^A_{\;\;BCD}n_An^Cg_\mu^Bg_\nu^D,
    \label{eq. projected weyl}
\end{equation}
where $C^A_{\;\;BCD}$ is the Weyl tensor. This non-local correction is traceless, \emph{i.e.}, $\mathcal{E}_\mu^\mu=0$.

Neves showed that a vacuum brane solution with $\tau_{\mu\nu}=\pi_{\mu\nu}=0$ can be obtained for a regular BH configuration~\cite{Neves:2021dqx}. By imposing the four-dimensional Hayward mass profile, the effective four-dimensional energy-momentum tensor takes the form\footnote{Note that $T^{(e)}_{\mu\nu}$ does not depend on the brane tension $\sigma$. In Ref.~\cite{Neves:2021dqx}, it is multiplied by a factor of $1/\kappa_5^2 k$, which at first glance suggests that $T^{(e)}_{\mu\nu}$ vanishes in the limit $\sigma \to \infty$. However, using the previously defined relation $\kappa_5^2 \propto 1/k$, the dependence on the brane tension is effectively factored out.}
\begin{equation}
    T^{(e)\;\nu}_{\qquad\mu}=\frac{\sigma}{2} - \frac{M_{BH}}{\kappa^2r^3}
    \begin{pmatrix}
        \mathcal{T}_1 & & & \\
        & \mathcal{T}_1 & & \\
        & & \mathcal{T}_2 & \\
        & & & \mathcal{T}_2
    \end{pmatrix},
    \label{eq. Tuv eff RBH}
\end{equation}
where
\begin{equation}
    \mathcal{T}_1(r)=\frac{6\left(\frac{\ell}{r}\right)^6+4\left(\frac{\ell}{r}\right)^3+1}{\left[1+\left(\frac{\ell}{r}\right)^3\right]^3}, \qquad\mathcal{T}_2(r)=\frac{6\left(\frac{\ell}{r}\right)^6-10\left(\frac{\ell}{r}\right)^3-1}{\left[1+\left(\frac{\ell}{r}\right)^3\right]^3}.
\end{equation}
The first term in Eq.~\eqref{eq. Tuv eff RBH} exactly cancels the brane tension contribution in Eq.~\eqref{eq. field eq brane}. It is therefore convenient to redefine
\begin{equation}
    T^{(H)}_{\mu\nu}=T^{(e)}_{\mu\nu}-\frac{\sigma}{2}g_{\mu\nu}.
\end{equation}

The non-local Weyl correction can be obtained by solving Eq.~\eqref{eq. projected weyl} using the metric in Eq.~\eqref{eq. metric line element 4d}, yielding
\begin{equation}
    \mathcal{E}_\mu^\nu=\frac{M_{BH}\left[1-2\left(\frac{\ell}{r}\right)^3\right]}{r^3\left[1+\left(\frac{\ell}{r}\right)^3\right]^3}
    \begin{pmatrix}
        -1 & & & \\
        & -1 & & \\
        & & 1 & \\
        & & & 1
    \end{pmatrix}.
    \label{eq. weyl component}
\end{equation}
It is common in the literature (see, e.g., Refs.~\cite{Castro:2014xza,Prasetyo:2017hrb} and references therein) to decompose the non-local Weyl tensor as
\begin{equation}
    \mathcal{E}_{\mu\nu} = -\frac{6}{\kappa^2\sigma}\left[U\left(u_\mu u_\nu + \frac{1}{3} h_{\mu\nu}\right) + P_{\mu\nu}\right],
    \label{eq. weyl form}
\end{equation}
with
\begin{equation}
    P_{\mu\nu} = P(r_\mu r_{\nu} - \frac{1}{3}h_{\mu\nu} ),
\end{equation}
where $P_{\mu\nu}$ is the anisotropic stress, $U$ is the bulk Weyl scalar, $u_\mu$ is a four-velocity field, $r_\mu$ is a unit radial vector, and $h_{\mu\nu}=g_{\mu\nu} + u_\mu u_\nu$. It is convenient to define
\begin{align}
    \tilde{U}=\frac{6}{\kappa^2\sigma}U,\qquad\tilde{P}=\frac{6}{\kappa^2\sigma}P,
\end{align}
together with the equation of state $\tilde{P}=\omega \tilde{U}$. From Eq.~\eqref{eq. weyl component}, one finds
\begin{equation}
    \tilde{U}=-\frac{M_{BH}\left[1-2\left(\frac{\ell}{r}\right)^3\right]}{r^3\left[1+\left(\frac{\ell}{r}\right)^3\right]^3},
\end{equation}
and $\tilde{P}=-2\tilde{U}$, corresponding to $\omega=-2$. In fact, this equation of state was also used in deriving the tidally charged BH solution~\cite{Dadhich:2000am}. We will adopt it throughout this work, as it allows us to determine the relevant physical quantities without introducing additional unknowns. Note that the explicit expressions for $U$ and $P$ are proportional to the brane tension $\sigma$ and thus become singular in the standard GR limit; we retain this feature as a consequence of the construction.

At this stage, it is useful to clarify a potential source of confusion. The final four-dimensional vacuum field equation no longer explicitly depends on the brane tension. Indeed, substituting $T^{(e)}{\mu\nu}$ and $\mathcal{E}{\mu\nu}$ into Eq.~\eqref{eq. field eq brane} yields
\begin{equation}
    G_\mu^\nu=\kappa^2
    \begin{pmatrix}
        -\varepsilon&&&\\&-\varepsilon&&\\&&q&\\&&&q
    \end{pmatrix},\label{eq. EFE hayward}
\end{equation}
where
\begin{equation}
    \varepsilon=M_{BH}\frac{\ell^3}{(r^3+\ell^3)^2},\qquad
    q=6M_{BH}\frac{\ell^3(2r^3-\ell^3)}{(r^3+\ell^3)^3}.
\end{equation}
This coincides with the field equations of a Hayward BH~\cite{Hayward:2005gi,Cadoni:2022chn}. In the limit $\ell \to 0$, one recovers the standard vacuum GR equation $G_{\mu\nu}=0$. This naturally raises the question: \emph{what roles do the brane tension $\sigma$ and higher-dimensional effects play in shaping the effective four-dimensional geometry?}

The key point is that, once the specific five-dimensional metric in Eq.~\eqref{eq. line element 5d} and the mass function in Eq.~\eqref{eq. mass hayward 5D} are imposed, the construction is effectively designed to produce a regular BH geometry on the brane. Consequently, the four-dimensional field equations\textemdash particularly for the Hayward BH\textemdash must take the form of Eq.~\eqref{eq. EFE hayward}. In this framework, the brane tension primarily influences the bulk rather than the brane directly: it controls how matter is distributed along the extra dimension. In particular, a larger $\sigma$ leads to stronger localization of bulk matter near the brane at $z=0$ (see Eq.~(28) of Ref.~\cite{Nakas:2020sey} and Eq.~(25) of Ref.~\cite{Neves:2021dqx}). Its effects on the brane become non-trivial only when a non-zero $\tau_{\mu\nu}$ is introduced, as will be discussed in the next section.

One may attempt to relate the regularization parameter $\ell$ to the anti-de Sitter length $\ell_{AdS}$, for instance by simply setting $\ell = \ell_{AdS}$. In that case, $\ell$ acquires an indirect dependence on the brane tension $\sigma$, and in the limit $\sigma \to \infty$ one obtains $\ell \to 0$, recovering the standard vacuum GR equation $G_{\mu\nu}=0$. This suggests a simple interpretation in which the regular BH geometry arises as a consequence of finite brane tension. However, such an identification introduces several physically significant issues, which will be discussed in Sec.~\ref{sec. significance}. Moreover, $\ell$ and $\ell_{AdS}$ carry fundamentally different physical meanings: $\ell$ is typically associated with a \emph{de Sitter} core scale of the BH, whereas $\ell_{AdS}$ corresponds to the \emph{anti-de Sitter} curvature scale of the five-dimensional bulk.

\section{Adding matter onto the brane}
\label{sec. matter on the brane}

Things become slightly more complicated when one adds matter onto the brane:

\textbf{(i) First}, we have a non-zero $\tau_{\mu\nu}$, which also implies a non-zero quadratic term $\pi_{\mu\nu}$. For our case, we employ the anisotropic fluid energy-momentum tensor given by $\tau_{\mu}^\nu=\text{diag}(-\epsilon,p,p_t,p_t)$. Solving the effective four-dimensional field equation in Eq.~\eqref{eq. field eq brane}, one obtains
\begin{align}
    e^{-\lambda}\left(\frac{\lambda'}{r} - \frac{1}{r^2}\right) + \frac{1}{r^2} &=-\kappa^2\epsilon^{eff} \label{eq. EFE effective 1},\\
    e^{-\lambda}\left(\frac{\nu'}{r}+\frac{1}{r^2}\right)-\frac{1}{r^2} &=\kappa^2p^{eff} \label{eq. EFE effective 2},\\
    e^{-\lambda}\left(\frac{1}{2}\nu'' - \frac{1}{4}\lambda'\nu'+\frac{1}{4}(\nu')^2+\frac{\nu'-\lambda'}{2r}\right)&=\kappa^2p_t^{eff}, \label{eq. EFE effective 3}
\end{align}
where the superscript $eff$ denotes the effective quantities, given by
\begin{align}
    \epsilon^{eff}=&\epsilon + \frac{\epsilon^2}{2\sigma}-\frac{(p-p_t)^2}{2\sigma} - T^{(H)\;0}_{\qquad 0} + \frac{\tilde{U}}{\kappa^2},
    \label{eq. epsilon eff}\\
    p^{eff}=&p - \frac{p^2}{2\sigma}+\frac{(\epsilon + p_t)^2}{2\sigma} + T^{(H)\;1}_{\qquad 1} + \frac{1}{3\kappa^2}\left(\tilde{U}+2\tilde{P}\right), \label{eq. p eff}\\
    p^{eff}_t=& p_t - \frac{p_t(p-\epsilon)}{2\sigma} + \frac{(\epsilon+p)^2}{2\sigma}-\frac{\epsilon p}{2\sigma} + T^{(H)\;2}_{\qquad2} + \frac{1}{3\kappa^2}\left(\tilde{U}-\tilde{P}\right). \label{eq. pt eff}
\end{align}
From the first field equation, Eq.~\eqref{eq. EFE effective 1}, one can directly obtain the differential equation for the mass profile $m(r)$ and the metric function $e^\lambda$:
\begin{equation}
    \frac{dm}{dr}=\frac{\kappa^2}{2}r^2\epsilon^{eff},\qquad e^{-\lambda}=1-\frac{2m(r)}{r}.
    \label{eq. dmdr elambda}
\end{equation}
It can be seen directly from Eq.~\eqref{eq. epsilon eff} that anisotropy ($p\neq p_t$) suppresses the mass contribution from the quadratic density term. In a near-classical regime, \emph{i.e.} $p_t\ll \sigma$, the quadratic term may not significantly affect the total effective energy density. However, we will see shortly that a special case may arise in which one potentially obtains a large transverse pressure close to horizon formation. 

\textbf{(ii) Second}, since our spacetime geometry is modified by the presence of matter, the non-local Weyl contribution $\mathcal{E}_{\mu\nu}$ will differ from that given in Eq.~\eqref{eq. weyl component}. In principle, the bulk energy-momentum contribution $T^{(H)}_{\mu\nu}$ may also be modified by the addition of matter. However, in this study, we assume that it remains unchanged, based on the argument that $T^{(H)}_{\mu\nu}$ originates purely from the bulk and is not affected by matter on the brane. This assumption also simplifies the analysis by reducing the number of unknowns. Therefore, the only quantity that changes is the Weyl contribution $\mathcal{E}_{\mu\nu}$, as previously mentioned, similar to the case of compact objects~\cite{Lugones:2017jak,Castro:2014xza,Prasetyo:2017hrb,Hladik:2011zz,GonzalezFelipe:2016edf,Murshid:2025ppz}.

One may note that the quantity $\nabla_\mu\tau^\mu_\nu$ is potentially non-zero, as Ref.~\cite{Neves:2021dqx} points out that $T^{(B)}_{MN}$ is non-diagonal, with an additional term $T^{(B)\;r}_{\qquad y}=e^{2\kappa|y|}T^{(B)\;y}_{\qquad r}$. It is known that~\cite{Maartens:2010ar}
\begin{equation}
    \nabla^{\mu}\tau_{\mu\nu}=-2T^{(B)}_{MN} n^Mg^N_\nu,
\end{equation}
which shows that $\nabla_\mu\tau^\mu_\nu\neq 0$ if $T_{ry}^{(B)}\neq0$. Fortunately, the quantity $T^{(B)\;r}_{\qquad y}$ vanishes on the brane ($y=0$), based on the metric in Eq.~\eqref{eq. line element 5d} and the field equation in Eq.~\eqref{eq. EFE 5D}. Hence, the matter on the brane remains conserved and satisfies $\nabla_\mu\tau^\mu_\nu=0$.

\subsection{Einstein cluster as the environment}
We employ the description of an Einstein cluster as the BH environment~\cite{Cardoso:2021wlq}. The energy-momentum tensor is written as $\tau_\mu^\nu=\text{diag}(-\epsilon_D,0,p_t,p_t)$, where $\epsilon_D$ and $p_t$ are the energy density and transverse pressure, respectively. We choose the Hernquist profile for the energy density distribution,
\begin{equation}
    \epsilon_D(r)=\epsilon_0\left(\frac{r}{a_0}\right)^{-\gamma}\left[1+\left(\frac{r}{a_0}\right)^\alpha\right]^{\frac{\gamma-\beta}{\alpha}},
    \label{eq. dark matter profile}
\end{equation}
with $(\alpha,\beta,\gamma)=(1,4,1)$. The parameters $a_0$ and $\epsilon_0$ are constants that determine the core size of the matter distribution and the core ``density'', respectively. This is known as the Hernquist DM profile~\cite{Hernquist:1990be}, and it is commonly used to describe DM phenomena in galaxies. Therefore, we will refer to the environment employed here as the \emph{DM distribution}. It should be emphasized that we do not attempt to explain DM-like effects, such as galaxy rotation curves sourced from the bulk, as in Refs.~\cite{Mak-Harko-can-2004,Bohmer-Harko-galactic-2007,Heydar-Fard:2007ahl}. Instead, we employ an astrophysical environment modeled by an Einstein cluster with a DM energy density profile surrounding the BH, which is widely referred to as a DM distribution or DM halo.

Several studies have pointed out that there is a sharp cut-off in the DM energy density near the BH horizon~\cite{Gondolo:1999ef,Sadeghian:2013laa}. The authors of Refs.~\cite{Cardoso:2021wlq,Speeney:2024mas,Figueiredo:2023gas} implement this cut-off by multiplying the energy density distribution by a cut-off factor,
\begin{equation}
\epsilon_D(r)\to\epsilon_D(r)\left(1-\frac{r_c}{r}\right),
\end{equation}
where $r_c$ is the cut-off radius at which the DM density vanishes. In Refs.~\cite{Cardoso:2021wlq,Figueiredo:2023gas}, $r_c$ is assumed to coincide with the BH horizon. However, it was found that the energy conditions are violated near the BH horizon~\cite{Datta:2023zmd}. To avoid this issue, Ref.~\cite{Speeney:2024mas} adopts $r_c=4M_{BH}$, following insights from the general relativistic treatment of DM annihilation in the vicinity of a BH~\cite{Sadeghian:2013laa}.

Let us now investigate the properties of this DM profile in the classical GR limit, \emph{i.e.} $\sigma\to\infty$, such that $\epsilon^{eff}(r)=\epsilon_D(r)$. The DM mass profile $m_{DGR}$ can be computed straightforwardly using Eq.~\eqref{eq. dmdr elambda}, yielding
\begin{equation}
    m_{DGR}(r)=M_{DGR}\frac{(r-r_c)^2}{(a_0+r)^2} ,
\end{equation}
where $M_{DGR}=\lim_{r\to\infty}m_{DGR}(r)$, and the quantity $M_{DGR}+M_{BH}$ corresponds to the ADM mass in GR. The parameter $M_{DGR}$ is related to $\epsilon_0$, $a_0$, and $r_c$ by
\begin{equation}
    M_{DGR}=\frac{\kappa^2}{4} \frac{\epsilon_0 a_0^4}{a_0+r_c}.
    \label{eq. MDGR}
\end{equation}
In the case of a non-negligible brane tension, the anisotropic term in Eq.~\eqref{eq. epsilon eff} reduces the effective energy density, thereby decreasing the total mass of the system. Consequently, in general, $\lim_{r\to\infty}m(r)\neq M_{BH}+M_{DGR}$.

\subsection{Solving the field equation}
\label{sec. solving field}

Now let us solve the remaining field equations while incorporating the DM distribution. The second field equation (Eq.~\eqref{eq. EFE effective 2}) gives the $g_{tt}$ metric coefficient
\begin{equation}
    \nu'=\frac{2m(r)+\kappa^2r^3p^{eff}}{r\left[r-2m(r)\right]}.
    \label{eq. dnudr}
\end{equation}
From $\nabla_\mu\tau^\mu_\nu=0$, we obtain
\begin{equation}
    p_t=\frac{1}{4}r\epsilon_D\nu'.
    \label{eq. transverse pressure}
\end{equation}
However, we see that the equations for $p_t$ and $\nu'$ are recursive: one needs $p_t$ to solve for $\nu'$, since $p^{eff}$ is a function of $p_t$, while $\nu'$ is also required to determine $p_t$. To resolve this, we employ an iterative method to determine $p_t$. We first take an initial approximation of the transverse pressure, $p_t^{(0)}$, in the form
\begin{equation}
    p_t^{(0)}=\frac{1}{4}r\epsilon_D\left(\nu'|_{p_t=0}\right),
\end{equation}
and perform the numerical calculation for $\tilde{U}(r)$ and $m(r)$. Once a solution is obtained, we iterate the calculation $j$ times using the corrected transverse pressure $p_t^{(i)}$,
\begin{equation}
    p_t^{(i)}=\frac{1}{4}r\epsilon_D\left(\nu'|_{p_t=p_t^{(i-1)}}\right),
\end{equation}
where $i$ runs from $1$ to $j$. However, from our tests, this method may lead to a divergent $p_t$ when $\sigma$ is sufficiently small and $\epsilon_D$ is large, which may arise because the final transverse pressure deviates significantly from the initial approximation. We also find that $\sim 4$ iterations are sufficient for the solutions to converge in typical cases.

Next, we investigate the continuity condition of the field equations. Since we know that $\nabla_\mu G^\mu_\nu=\nabla_\mu\tau^\mu_\nu=0$, one requires
\begin{equation}
    \nabla_\mu\left(\kappa^2T^{(H)\;\mu}_{\qquad\nu} + \frac{6\kappa^2}{\sigma}\pi^\mu_\nu - \mathcal{E}^\mu_\nu\right)=0.
    \label{eq. conservation all}
\end{equation}
The contribution from the regular BH effective tensor $\nabla_\mu T^{(H)\;\mu}_{\qquad\nu}$ is
\begin{align}
    \nabla_\mu \kappa^2 T^{(H)\;\mu}_{\qquad r}=& \frac{d\mathcal{T}_1}{dr}+2\frac{\mathcal{T}_1-\mathcal{T}_2}{r}\notag\\
    =& -\frac{Mr^2\left(r^6+14r^3\ell^3-14\ell^6\right)}{\left(r^3+\ell^3\right)^4},
\end{align}
while the quadratic term is more cumbersome, given by
\begin{equation}
    \nabla_\mu\pi^\mu_r=\frac{1}{12r}\left[-p^2\left(2+r\nu'\right)\right.\notag \left.+p\mathcal{A}(r)+\mathcal{B}(r)\right],
    \label{eq. nabla pi}
\end{equation}
where
\begin{align}
    \mathcal{A}(r)=&p_t-\epsilon_D-2rp'+rp_t\nu',\notag\\
    \mathcal{B}(r)=&\left(p_t+\epsilon_D\right)\left[p_t+2r\left(p_t'+\epsilon_D'\right)+r\epsilon_D\nu'\right].
\end{align}
However, for the vanishing radial pressure DM used in our model, the surviving term reads
\begin{equation}
    \nabla_\mu\pi^\mu_r=\frac{\left(p_t+\epsilon_D\right)}{12r}\left[p_t+2r\left(p_t'+\epsilon'\right)+r\epsilon_D\nu'\right].
\end{equation}
Finally, using Eq.~\eqref{eq. weyl form} for the non-local correction, we obtain
\begin{align}
    \nabla_\mu\mathcal{E}^\mu_r = &-\frac{1}{3}\left[(\tilde{U}+2\tilde{P})'+2(2\tilde{U}+\tilde{P})\nu'+\frac{6}{r}\tilde{P}\right]\notag\\
    =& \tilde{U}'+\frac{4}{r}\tilde{U}.\qquad(\tilde{P}=-2\tilde{U})
\end{align}
From this, we obtain a differential equation for $\tilde{U}$:
\begin{equation}
    \tilde{U}'+ \frac{4}{r}\tilde{U}=\frac{\kappa^2}{2\sigma r} \left(p_t+\epsilon_D\right)\left[p_t+2r\left(p_t'+\epsilon_D'\right)+r\epsilon_D\nu'\right]-\frac{M_{BH}r^2\left(r^6+14r^3\ell^3-14\ell^6\right)}{\left(r^3+\ell^3\right)^4}.
    \label{eq. differential U}
\end{equation}
The boundary condition for Eq.~\eqref{eq. differential U} is $\tilde{U}(r)|_{r=0}=0$, following the known form of $\tilde{U}(r)$ in the absence of matter on the brane, \emph{i.e.} Eqs.~\eqref{eq. weyl component} and~\eqref{eq. weyl form}. This differs from neutron star studies in braneworld scenarios, where the boundary condition is typically $\tilde{U}(r)|_{r=0}=\tilde{U}_c$, with $\tilde{U}_c$ chosen such that $\tilde{U}$ vanishes at the stellar surface~\cite{Lugones:2017jak,Castro:2014xza,Germani:2001du,Prasetyo:2017hrb,Hladik:2011zz,GonzalezFelipe:2016edf,Murshid:2025ppz}.

To verify this equation, one can solve Eq.~\eqref{eq. differential U} in the brane vacuum case ($\epsilon_D=p_t=0$), yielding
\begin{equation}
    \tilde{U}=-\frac{M_{BH}\left[1-2\left(\frac{\ell}{r}\right)^3\right]}{r^3\left[1+\left(\frac{\ell}{r}\right)^3\right]^3} + \frac{Q}{r^4},
\end{equation}
which agrees with Eq.~\eqref{eq. weyl component} with an additional $Q/r^4$ term. The $Q$ itself is an integration constant, which, in fact, is the BH tidal charge~\cite{Dadhich:2000am}. Hence, in the case of the pure Hayward regular BH spacetime, there is no tidal charge arising from the bulk, \emph{i.e.} $Q=0$.

The above equations reveal a peculiar behavior of the DM distribution and its effect on the spacetime geometry. From Eqs.~\eqref{eq. transverse pressure} and~\eqref{eq. dnudr}, we see that the transverse pressure can become very large near horizon formation, \emph{i.e.} as $2m(r)\to r$. From Eq.~\eqref{eq. epsilon eff}, a large transverse pressure can significantly reduce the effective energy density, or even decrease the total mass if the quadratic transverse pressure term dominates over the other contributions. Furthermore, the condition $2m(r)\to r$ also induces a strong variation in $\tilde{U}$, as implied by Eq.~\eqref{eq. differential U}, which may either suppress or enhance the contribution of $p_t$ to the effective energy density. If these effects lead to a vanishing effective energy density ($\epsilon^{eff}\to 0$) at the critical point $r\to 2m(r)$, horizon formation due to the DM distribution may be avoided.

A consequence of this is that, with finite brane tension, one may have an arbitrarily dense DM distribution without ever forming a horizon. This is in line with the findings in Ref.~\cite{Lugones:2017jak}, where finite brane tension allows a neutron star to attain arbitrarily large masses. It is also worth noting that this occurrence offers an interesting perspective for further studies on the impact of finite brane tension on compact stars with anisotropic pressure.

\subsection{Significance of the braneworld}
\label{sec. significance}
Before solving the equations, one can predict for which DM configurations these braneworld effects take place. It is useful to normalize all quantities with respect to the BH mass $M_{BH}$,
\begin{gather}
    \tilde{p}_t=p_tM_{BH}^2,\quad \tilde{\epsilon}_D=\epsilon_DM_{BH}^2, \quad \tilde{\sigma}=\sigma M_{BH}^2,\notag\\ \tilde{\ell}=\frac{\ell}{M_{BH}},\quad \tilde{a}_0=\frac{a_0}{M_{BH}}, \quad \tilde{r}_c=\frac{r_c}{M_{BH}},
\end{gather}
which provides a proper scaling to obtain dimensionless forms of each quantity. From Eq.~\eqref{eq. field eq brane}, the braneworld modifications arise from the quadratic high-energy term, \emph{i.e.} $\sigma^{-1}\pi_{\mu\nu}$. Therefore, we expect significant modifications when either $\tilde{\epsilon}_{D}^2$, $\tilde{p}_t^2$, or $\tilde{\sigma}^{-1}$ attains non-negligible values.

\subsubsection{Determining the relevant scenario}
The profile for $\epsilon_D$ is given in Eq.~\eqref{eq. dark matter profile}, where the maximum value of $\epsilon_D$ is primarily determined by $\epsilon_0$ (or $M_{DGR}$, c.f. Eq.~\eqref{eq. MDGR}). Within classical GR, setting a large value of $\epsilon_0$ may lead to horizon formation, especially for small $a_0$. There exists a maximum limit $\epsilon_0^{M}$ for a given $a_0$ such that the DM distribution does not form an additional horizon, which can be obtained by finding a suitable $\epsilon_0$ that satisfies
\begin{equation}
e^{-\lambda}=\partial_re^{-\lambda}=0
\end{equation}
outside the BH ($r>r_H$). Using this $\epsilon_0^{M}$, we compute the maximum of $\tilde{\epsilon}_D$ by solving $\partial_r\tilde{\epsilon}_D=0$. We find that the maximum value of $\tilde{\epsilon}_D$ is always less than unity, regardless of the DM core size $\tilde{a}_0$ or the cut-off radius $\tilde{r}_c$. Hence, the DM energy density model employed in this study does not provide a sufficiently large contribution to distinguish between braneworld and classical GR effects for large $\tilde{\sigma}$.

Another quantity we can consider is $\tilde{p}_t$. From Eq.~\eqref{eq. transverse pressure}, large value of $\tilde{p}_t$ can be achieved in high local compactness $m(r)/r$, particularly as it approaches horizon formation, \emph{i.e.} $2m(r)\to r$. However, such conditions are not expected for DM distributions in galaxies\textemdash especially those with extended cores\textemdash as they would render the central region extremely redshifted: Assuming an asymptotically flat spacetime with $\nu(r)|_{r\to\infty}=0$, Eq.~\eqref{eq. dnudr} implies that $\nu$ increases outward (or decreases inward). As $2m(r)\to r$, the $\nu$ coefficient would approaches zero toward the center, unless a region with negative effective radial pressure exists, satisfying $-p^{eff}>2m(r)/\kappa^2r^3$, which could increase $\nu$ inward and prevent excessive redshift.

From the above analysis, two possible scenarios in which braneworld effects may become relevant are: (i) the DM distribution is ultracompact, or (ii) the BH mass is small\textemdash for which, to our knowledge, there is no observational evidence of such situations. Consequently, we conclude that there will be no significant signature of braneworld modifications for DM distributions surrounding BHs with configurations expected by current observations. 

For reference, let us consider the brane tension in the range $10^{36}-10^{38}\;\text{dyn/cm}^2$~\cite{GonzalezFelipe:2016edf,Murshid:2025ppz}. If one requires $\tilde{\sigma}\ll1$\textemdash for instance, $\tilde{\sigma}\approx0.001$\textemdash the BH mass must lie in the range $\sim 1.0-0.1\,M_\odot$. On the other hand, for a BH with the mass of Sgr A* (approximately $\sim 4.3\times10^{6}\,M_\odot$), the normalized brane tension becomes $\tilde{\sigma}\approx 3\times10^{10}$. Therefore, braneworld signatures are much more likely to be observable around stellar-mass BHs surrounded by relatively compact DM halos, and are practically unobservable at the scale of supermassive BHs in galaxies.

\subsubsection{Connecting $\sigma$ and $\ell$?}

Both regular BH solutions and the RS-II framework introduce a characteristic length scale, denoted by $\ell$ and $\ell_{AdS}$, respectively. In both cases, the standard GR solution is recovered in the limit $\ell = \ell_{AdS} = 0$. Since $\ell_{AdS} = 1/k$ is related to the brane tension through $\sigma = 6(\kappa^2 \ell_{AdS}^2)^{-1}$, one might be tempted to identify $\ell \equiv \ell_{AdS}$, thereby making $\ell$ dependent on the brane tension. In this picture, taking the limit $\ell \to 0$ corresponds to $\sigma \to \infty$, and standard vacuum GR is recovered.

However, several complications arise from this identification. A BH solution exists only if the dimensionless parameter $\tilde{\ell}$ does not exceed a critical value $\tilde{\ell}_c$. For the Hayward BH, this critical value is $\tilde{\ell}_c = 2\sqrt[3]{4}/3$~\cite{Cadoni:2022chn}, which implies a lower bound on the dimensionless brane tension, $\tilde{\sigma} \gtrsim 0.21$. As will be shown in the next section, such values lead to a negligible influence of braneworld corrections on both the matter distribution and the spacetime geometry in realistic astrophysical scenarios.

Additional issues arise from observational constraints based on BH shadows. Measurements of the Sgr A* shadow radius indicate that it can be described by a Hayward BH over essentially the entire range $0 \leq \tilde{\ell} \leq \tilde{\ell}_c$~\cite{Vagnozzi:2022moj}. This translates into a minimum brane tension of $\sigma \gtrsim 6.75 \times 10^{24}\,\text{dyn/cm}^2$, which is several orders of magnitude lower than bounds inferred from neutron star observations~\cite{GonzalezFelipe:2016edf,Murshid:2025ppz} (see also Ref.~\cite{Germani:2001du}). Although one may consider alternative regular BH models whose parameters are constrained before reaching their critical values, these parameters are typically still of the same order as the BH mass~\cite{Vagnozzi:2022moj}, again leading to brane tension bounds that are significantly weaker than those derived from neutron stars.

The situation becomes even less reliable when DM is included in the shadow analysis. DM distributions around a BH tend to increase the shadow radius~\cite{Xavier:2023exm,Fauzi:2025yse,Macedo:2024qky,Fernandes:2025osu}, whereas the regularization parameter $\ell$ generally decreases it~\cite{Vagnozzi:2022moj}. The interplay between DM effects and finite brane tension therefore introduces a degeneracy in the parameter space, making meaningful constraints difficult to obtain. For instance, since the Schwarzschild BH shadow radius already lies close to the upper bound of the Sgr A* observations, adding a sufficiently compact DM distribution (e.g., with $a_0 = 10^3 M_{BH}$ and $M_{DGR} \gtrsim 67 M_{BH}$~\cite{Fauzi:2025yse}, albeit unrealistic for Sgr A*) would increase the shadow size beyond the allowed range. This effect could then be compensated by increasing $\ell$ (or equivalently decreasing $\sigma$), bringing the shadow back within observational bounds. Such a degeneracy may lead to an artificial upper bound on the brane tension that lacks clear physical significance.

For these reasons, we do not adopt the identification $\ell \equiv \ell_{AdS}$ in the remainder of this work. Instead, we treat $\ell$ and $\ell_{AdS}$ as independent parameters with distinct physical interpretations.

\section{Resulting braneworld modification}
\label{sec. result braneworld modification}

For a concise study, we choose to employ the Schwarzschild BH ($\ell\to0$) in our calculations. We find that, even in the extremal configuration of a regular BH with $\tilde{\ell}=2\sqrt[3]{4}/3$, the spacetime geometry outside the BH and the DM characteristics barely deviate from those of the Schwarzschild BH. Therefore, it is sufficient to consider only the singular BH model to study the influence of the braneworld on the DM distribution.

As previously mentioned, the impact of finite brane tension becomes significant for small values of the dimensionless brane tension relative to the BH mass. We specifically choose $\tilde{\sigma}\in\{0.01,0.005,0.001\}$ in our calculations, which, assuming a brane tension value of $\sigma=10^{37}\;\text{dyn/cm}^2$~\cite{GonzalezFelipe:2016edf,Murshid:2025ppz}, corresponds to $M_{BH}/M_{\odot}\in\{0.74,0.53,0.23\}$. These masses lie below the Chandrasekhar limit, and BHs with such masses cannot be formed from standard stellar collapse. However, recent studies suggest that such BHs may originate from Hawking stars~\cite{Santarelli:2024uqx}. In such models, a small primordial BH in environment may be captured by a star and gradually accrete stellar material from within, eventually leading to the formation of a sub-stellar BH remnant. This mechanism provides a possible astrophysical motivation for considering the low-mass BH studied here, together with their surrounding matter distributions.

We compare several energy density configurations of the DM distribution in the following discussion. In other words, we aim to examine how the brane tension affects the same matter content, \emph{i.e.} the same energy density distribution. We choose $M_{DGR}/M_{BH}\in\{10,20,30\}$ with $a_0/M_{BH}\in\{10,20,40\}$; the corresponding $\epsilon_0$ values follow from Eq.~\eqref{eq. MDGR}. These values are chosen such that the DM distribution is neither too dilute nor too compact for most parameter combinations, allowing us to clearly observe the impact of finite brane tension.

\subsection{ADM mass discrepancy}

\begin{figure}[htbp!]
    \centering
    \includegraphics[width=0.8\textwidth]{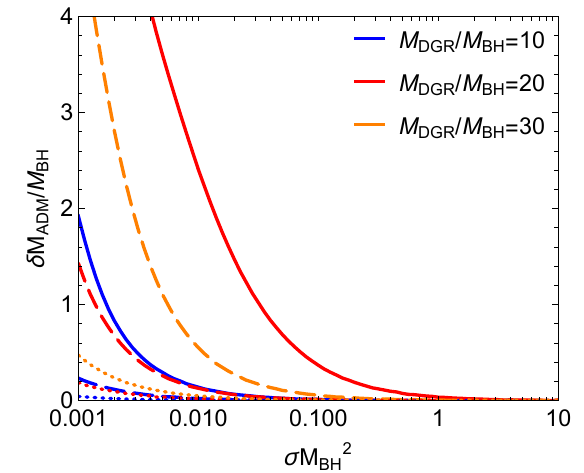}
    \caption{ADM mass discrepancy between the finite brane tension case and classical GR. The solid, dashed, and dotted lines represent configurations with $a_0/M_{BH}=10$, $a_0/M_{BH}=20$, and $a_0/M_{BH}=40$, respectively. It is shown that, for the same matter content, the total ADM mass decreases as the brane tension becomes smaller.}
    \label{fig. delta adm}
\end{figure}

From the analysis of Eq.~\eqref{eq. epsilon eff}, a large anisotropy in the presence of finite brane tension reduces the effective mass contribution of the DM, leading to a smaller total ADM mass compared to the classical GR case. We write discrepancy between the ADM mass in the braneworld and classical GR cases as
\begin{equation}
    \delta M_{ADM}=(M_{BH}+M_{DGR})-\lim_{r\to\infty}m(r), 
\end{equation}
and resulting discrepancy as a function of the brane tension is shown in Fig.~\ref{fig. delta adm}.

It is evident that the ADM mass decreases as the brane tension becomes smaller. This reduction is more pronounced for more compact DM distributions, as a smaller core size $a_0$ with the same $M_{DGR}$ leads to a larger decrease in the ADM mass. As previously discussed, this behavior arises because finite brane tension has a stronger impact in regions of high local compactness, which enhances the transverse pressure and consequently reduces the effective energy density. The implication is that, for the same matter content surrounding the BH, finite brane tension leads to a weaker gravitational influence on distant objects.

\begin{figure}[htbp!]
    \centering
    \includegraphics[width=1\textwidth]{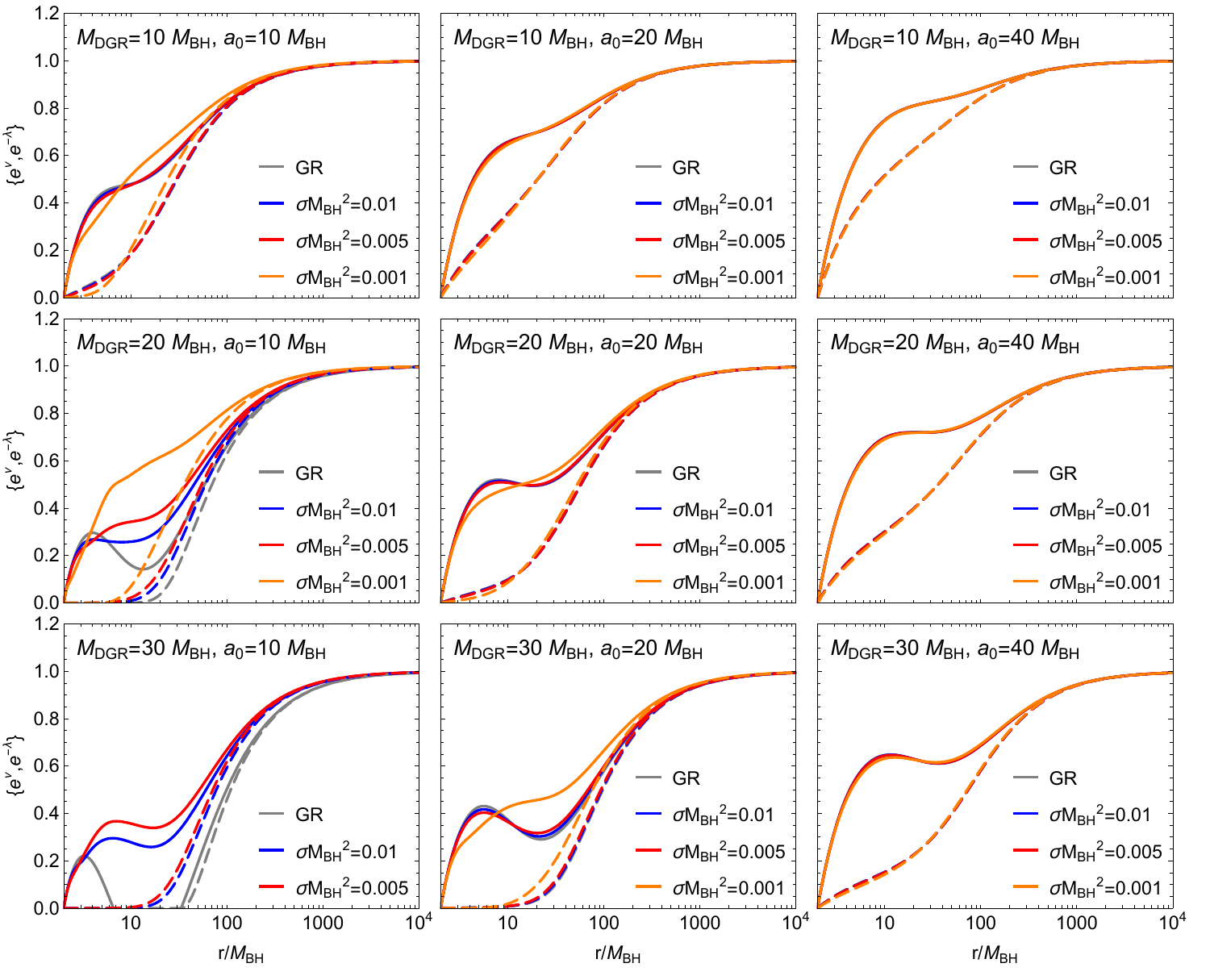}
    \caption{Time (solid lines) and radial (dashed lines) components of the spacetime geometry for all combinations of $M_{DGR}/M_{BH}$ and $a_0/M_{BH}$ considered in this study. The transverse pressure\textemdash and consequently the spacetime geometry\textemdash in the bottom-left panel does not converge for $\sigma M_{BH}^2=0.001$. For higher values of the brane tension, the solutions yield $e^{-\lambda}>0$ everywhere outside the BH horizon, in contrast to the classical GR case.}
    \label{fig. gtt grr}
\end{figure}

\subsection{Spacetime geometry}
\label{sec. spacetime geometry}

One of the key aspects of our study is the impact of brane tension on the spacetime geometry arising from the DM distribution. In Fig.~\ref{fig. gtt grr}, we show the time and radial components of the spacetime metric outside the BH for all considered configurations. A direct effect is the flattening of $g_{rr}^{-1}$ for smaller brane tension, which clearly originates from the reduction of the mass function $m(r)$ due to finite brane tension, although the detailed flattening pattern is not entirely uniform. For $g_{tt}$, however, the effect is less straightforward. A noticeable difference across all configurations is that larger brane tension leads to higher values of $-g_{tt}$ at scales larger than $a_0$. Closer to the BH, its behavior depends on the compactness of the DM distribution: more compact configurations with $a_0/M_{DGR}<1$ tend to exhibit higher $-g_{tt}$ at $r<a_0$ for smaller brane tension, while the opposite behavior is observed for $a_0/M_{DGR}\gtrsim1$.

The most compact configuration, with $a_0/M_{BH}=10$ and $M_{DGR}/M_{BH}=30$, confirms our analysis from the previous section. In classical GR, this configuration gives rise to horizon formation induced by the DM, which introduces curvature singularities at the additional horizons~\cite{Cardoso:2021wlq}. Finite brane tension, however, prevents the formation of such horizons, as both metric components remain non-vanishing everywhere outside the BH horizon at $r=2M_{BH}$. Nevertheless, the curvature properties in the finite brane tension case remain unclear, since a regular metric does not, in general, guarantee spacetime regularity~\cite{Rodrigues:2023fps}. We do not pursue this direction further and leave it for future work.

Although we have verified that horizon formation can be avoided, an arbitrarily dense DM distribution may lead to a violation of the causal limit, as the speed of sound in the matter distribution must remain below the speed of light. We will discuss this issue further in the following section.

\begin{figure}[htbp!]
    \centering
    \includegraphics[width=1\textwidth]{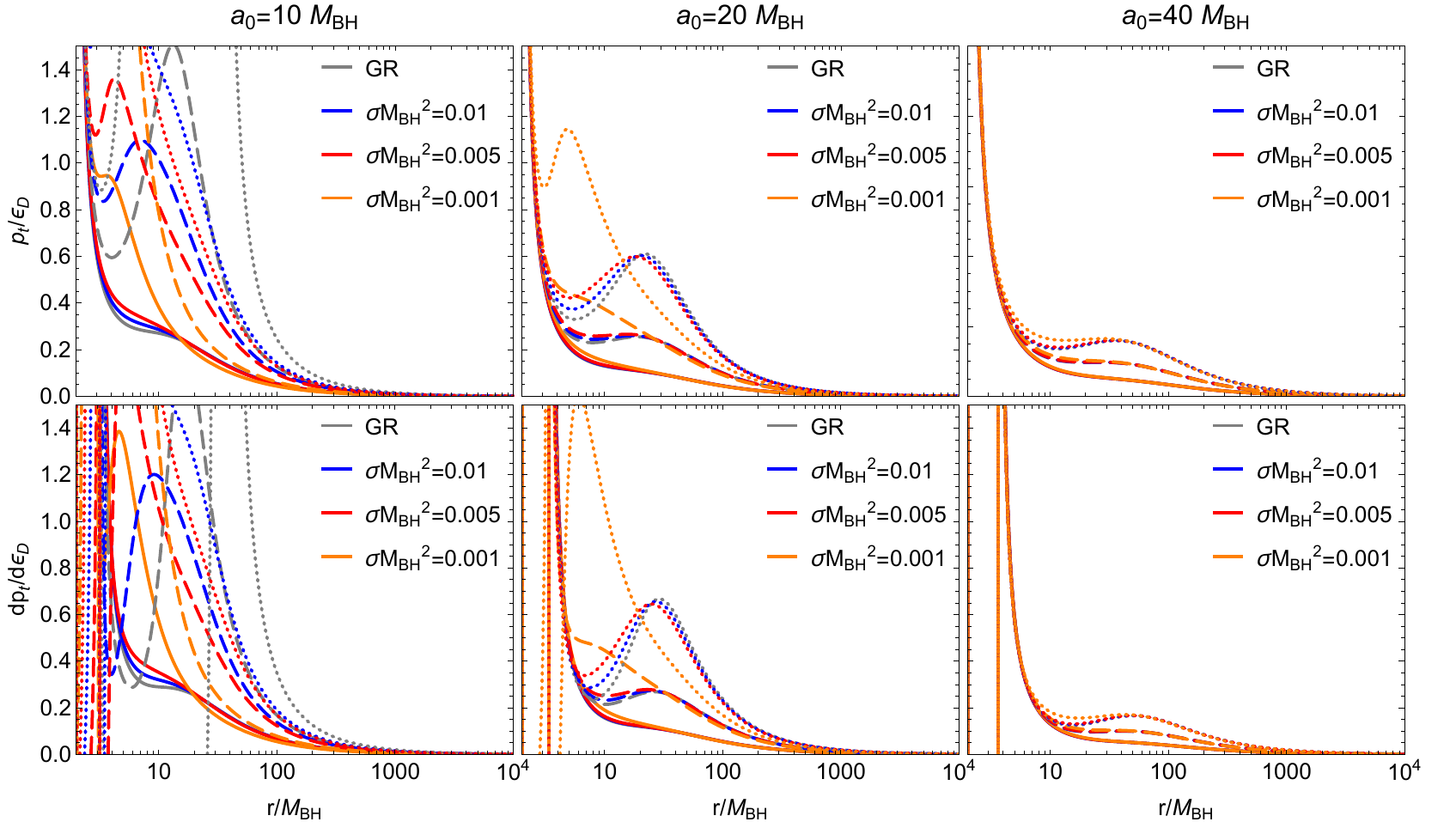}
    \caption{(Top) Dominant energy condition and (Bottom) squared speed of sound of the DM distribution. The solid, dashed, and dotted lines represent configurations with $M_{DGR}/M_{BH}=10$, $M_{DGR}/M_{BH}=20$, and $M_{DGR}/M_{BH}=30$, respectively. In all cases, more compact configurations produce higher local maxima for both the DEC and the speed of sound, and the brane tension generally modifies these behaviors.}
    \label{fig. ec cs}
\end{figure}

\subsection{Energy conditions and speed of sound}

As a consequence of the distinct transverse pressure of the DM, the satisfaction of the energy conditions differs from that in classical GR. The three standard energy conditions include the strong energy condition (SEC), weak energy condition (WEC), and dominant energy condition (DEC)~\cite{Hawking:1973uf}. Since we find that the transverse pressure is always positive for all configurations considered, and noting the vanishing radial pressure, both the SEC and WEC are trivially satisfied.

The DEC, on the other hand, imposes a stricter requirement, namely $|p_t|<\epsilon_D$~\cite{Hawking:1973uf}. It has been shown that the DEC is violated near the horizon for an Einstein cluster DM distribution in classical GR~\cite{Datta:2023zmd}. The DEC is also closely related to the speed of sound $c_s$, which can be written as
\begin{equation}
    c_s^2=\frac{dp_t}{d\epsilon_D}=\frac{dp_t}{dr}\left(\frac{d\epsilon_D}{dr}\right)^{-1},
\end{equation}
for the transverse direction. The causal limit requires that the speed of sound remains below the speed of light, \emph{i.e.} $c_s<1$. We present the relevant quantity characterizing the DEC, \emph{i.e.} $p_t/\epsilon_D$, along with the squared speed of sound for our chosen configurations in Fig.~\ref{fig. ec cs}.

The results indicate that finite brane tension shifts the local maxima of $p_t/\epsilon_D$ inward. In certain regions of parameter space, specifically for the configuration with $a_0/M_{BH}=20$ and $M_{DGR}/M_{BH}=20$, sufficiently small brane tension leads to violations of both the DEC and the causal limit. The results also suggest that there may exist an optimal range of brane tension that partially alleviates both the DEC violation and the excessive transverse sound speed, as the peak values initially decrease with decreasing brane tension but increase again beyond a certain point. Furthermore, ultracompact DM configurations with small cores\textemdash particularly those with $M_{DGR}>10M_{BH}$ and $a_0=10M_{BH}$\textemdash generally violate both the DEC and the causal limit. Hence, we confirm that, although horizon formation can be avoided with finite brane tension, violations of the causal limit may still occur.

\section{Photon spheres, shadow, and Einstein ring radius}
\label{sec. observables}

\begin{figure*}[htbp!]
    \centering
    \includegraphics[width=1\textwidth]{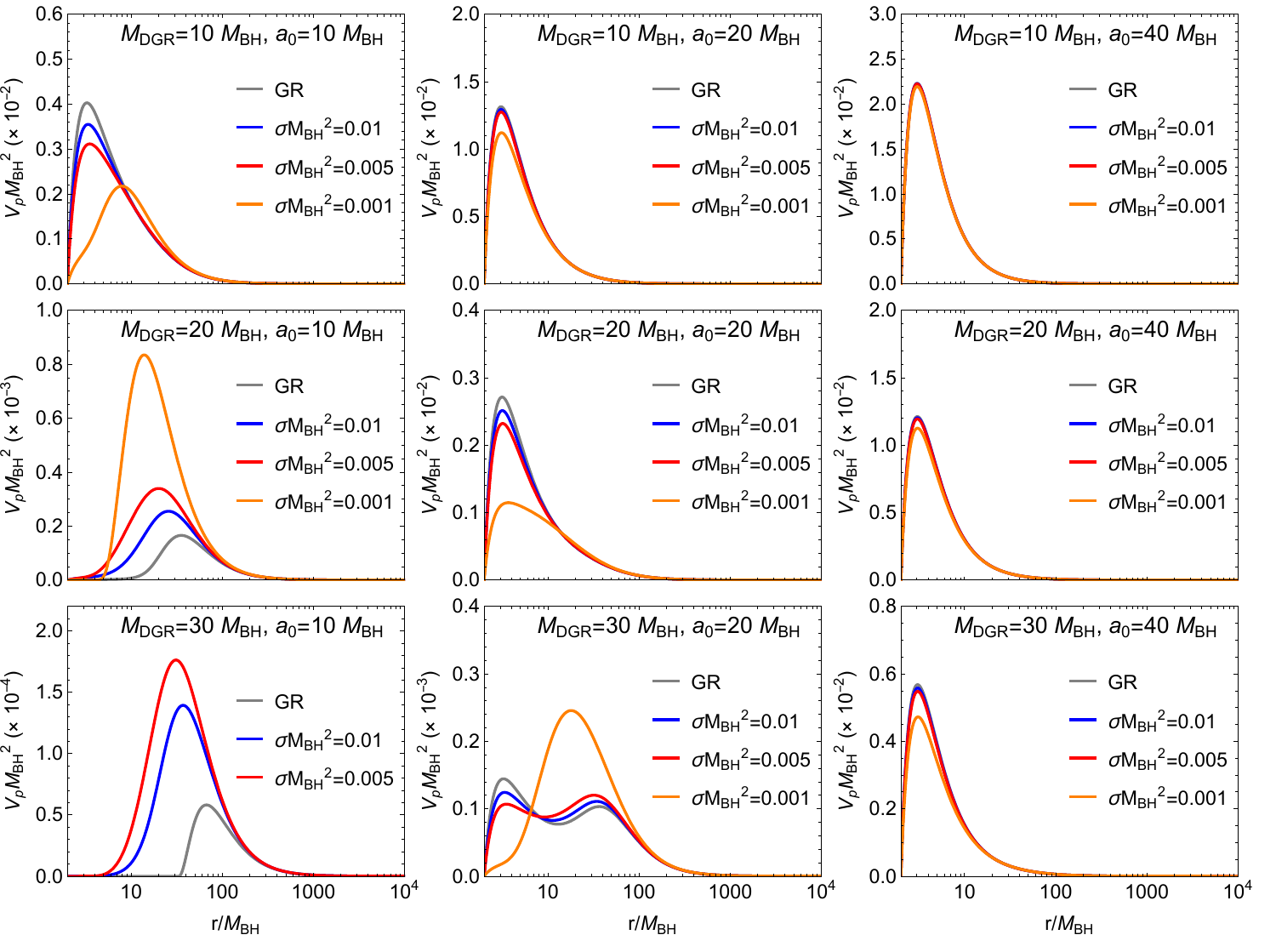}
    \caption{Effective potential for photon geodesics around the BH and the DM distribution. The solid, dashed, and dotted lines represent configurations with $a_0/M_{BH}=10$, $a_0/M_{BH}=20$, and $a_0/M_{BH}=40$, respectively. In general, three distinct types of potentials are observed: one peak near the BH horizon, one peak near the scale of $a_0$, and a configuration with two peaks appearing simultaneously. Note the scaling factor on the vertical axis label.}
    \label{fig. photon veff}
\end{figure*}

We saw in the previous section that the presence of DM results in different spacetime characteristics for various configurations in the braneworld scenario. The modified spacetime must therefore alter particle dynamics around the BH. It should be noted that the higher-dimensional spacetime does not explicitly modify the particle geodesics; instead, the modification is implicitly encoded in the effective four-dimensional spacetime geometry satisfying Eq.~\eqref{eq. field eq brane}. Therefore, we solve the geodesic equation in 4 dimensions,
\begin{equation}
    \frac{d^2x^\mu}{d\xi^2}+\Gamma_{\alpha\beta}^\mu \frac{dx^\alpha}{d\xi}\frac{dx^\beta}{d\xi}=0,
\end{equation}
where $\xi$ is an affine parameter, $x^\mu=(t,r,\theta,\phi)$, and $\Gamma_{\alpha\beta}^\mu$ denotes the Christoffel symbols.

In the following discussion, we examine several quantities related to photon dynamics around the system, specifically the photon sphere radius $r_{ps}$, the BH shadow radius $R_{sh}$, and the Einstein ring radius $\Theta_{ER}$. These quantities are clearly illustrated in the supplementary material of Ref.~\cite{Fauzi:2025yse}, and we adopt the same notation used in that study. We also provide the numerical values of the direct observable quantities ($R_{sh}$ and $\Theta_{ER}$) in the classical GR case in Table~\ref{tab. Rsh ER GR}.

\begin{table}[htbp!]
\centering
\setlength{\tabcolsep}{7pt}
\begin{tabular}{cccc}
\hline\hline
$a_0/M_{BH}$ & $M_{DGR}/M_{BH}$ & $R_{sh}^{(GR)}/M_{BH}$ & $\Theta^{(GR)}_{ER}\;[\mu\text{as}]$\\
\hline
10 & 10 & 15.76 & 4469.92 \\
& 20 & 77.70 & 6418.53 \\
20 & 10 & 8.72 & 4308.58 \\
& 20 & 19.19 & 6239.55\\
& 30 & 83.22 & 7823.38\\
40 & 10 & 6.69 & 3997.92 \\
& 20 & 9.08 & 5888.72\\
& 30 & 13.27 & 7451.53\\
\hline
\end{tabular}
\caption{Numerical values of the classical GR shadow radius $R_{sh}^{(GR)}$ and the Einstein ring angular radius $\Theta^{(GR)}_{ER}$ for all configurations considered in this study. The Einstein ring radius corresponds to a lensing setup with $D_{OL}=10^{11}M_{BH}$ and $D_{LS}=10^{3}M_{BH}$. Discussions of both quantities can be found in Sec.~\ref{sec. shadow radius} and Sec.~\ref{sec. einstein ring radius}.}
\label{tab. Rsh ER GR}
\end{table}

\subsection{Null geodesics and photon sphere radius}

The 4-velocity condition for null geodesics reads
\begin{equation}
g_{tt}\dot{t}^2+g_{rr}\dot{r}^2+r^2\dot{\theta}^2+r^2\sin^2\theta\dot{\phi}^2=0,
\label{eq. 4 velocity null}
\end{equation}
where the dot denotes derivatives with respect to the affine parameter. Solving the geodesic equation confined to the equatorial plane ($\theta=\pi/2$, $\dot{\theta}=0$), one finds two conserved quantities, $L=r^2\dot{\phi}$ and $E=-g_{tt}\dot{t}$, corresponding to the angular momentum and energy of the particle, respectively. Rearranging Eq.~\eqref{eq. 4 velocity null} and substituting these conserved quantities yields
\begin{equation}
-g_{tt}g_{rr}\left(\frac{dr}{d\tau}\right)^2+V_p(r)L^2=E^2,\qquad V_p(r)=\frac{-g_{tt}}{r^2},
\label{eq. geodesic T+V=E}
\end{equation}
where $V_p(r)$ is the photon effective potential. Applying the chain rule, one obtains
\begin{equation}
	\frac{d\phi}{dr} = \frac{1}{r^2} \left[\frac{-g_{tt}g_{rr}}{b^{-2} - V_p(r)}\right]^{1/2},
	\label{eq. radial geodesic equation} 
\end{equation}
where $b=L/E$ is the impact parameter. We show the photon effective potential in Fig.~\ref{fig. photon veff}.

\begin{figure}[htbp!]
    \centering
    \includegraphics[width=1\textwidth]{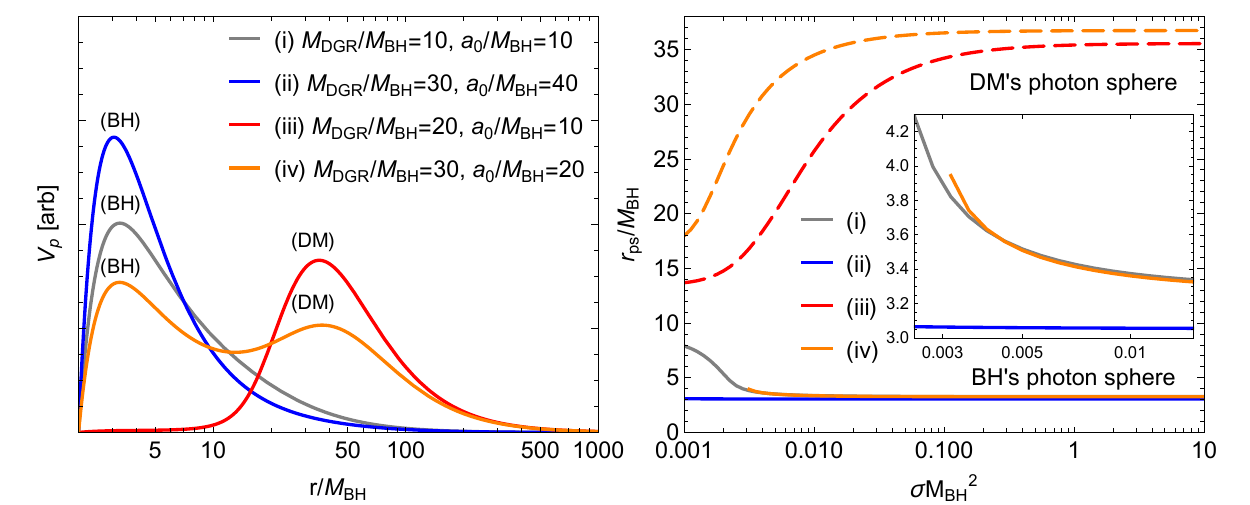}
    \caption{(Left) Photon effective potential in the classical GR case, illustrating the photon sphere location and its classification. (Right) Photon sphere radius for the chosen representative configurations. The solid and dashed lines in the right panel represent the photon spheres produced by the BH and the DM distribution, respectively. In general, the DM photon sphere decreases with smaller brane tension, while the BH photon sphere behaves in the opposite manner.}
    \label{fig. rps vp}
\end{figure}

Within the null geodesic framework, one can determine the photon sphere radius $r_{ps}$, defined as the radius at which photons orbit indefinitely around the center ($r=0$). This is obtained from the conditions $dr/d\tau=d^2r/d\tau^2=0$ at $r=r_{ps}$, which lead to
\begin{equation}
    \left.\frac{dV_p}{dr}\right|_{r=r_{ps}}=0.
\end{equation}
Thus, the existence of photon spheres can be directly inferred from the extrema of the effective potential. We show the photon effective potential for several representative cases in Fig.~\ref{fig. rps vp}.

We classify photon spheres into two categories: those produced by the BH geometry and those arising from the DM distribution. The latter can only appear when the DM is sufficiently compact and may coexist with the BH photon sphere. However, for extremely compact DM configurations, the BH photon sphere may be suppressed, leaving a photon sphere purely induced by the DM. The left panel of Fig.~\ref{fig. rps vp} illustrates this scenario in the classical GR case.

In the right panel of Fig.~\ref{fig. rps vp}, we observe modifications to both photon sphere radii due to finite brane tension. The BH and DM photon spheres exhibit opposite behavior: as the brane tension decreases, the DM photon sphere radius decreases, while the BH photon sphere radius increases. In cases where both photon spheres coexist, sufficiently small brane tension shifts the DM photon sphere inward, eventually leading to the disappearance of the BH photon sphere.

\subsection{Shadow radius}
\label{sec. shadow radius}
The BH shadow radius can be obtained from the critical impact parameter $b_c$ associated with the photon sphere, specifically the innermost photon sphere accessible to an observer at infinity. By requiring the radial motion in Eq.~\eqref{eq. radial geodesic equation} to vanish at the photon sphere, one finds
\begin{equation}
	b_c=\left[V_p(r_{ps})\right]^{-1/2}.
\end{equation}
Assuming the observer is located far from the BH, the shadow radius is given by $R_{sh}=b_c$~\cite{Perlick:2021aok}. This relation indicates that the shadow radius is inversely proportional to the maximum value of the photon effective potential.

\begin{figure}[htbp!]
    \centering
    \includegraphics[width=0.8\textwidth]{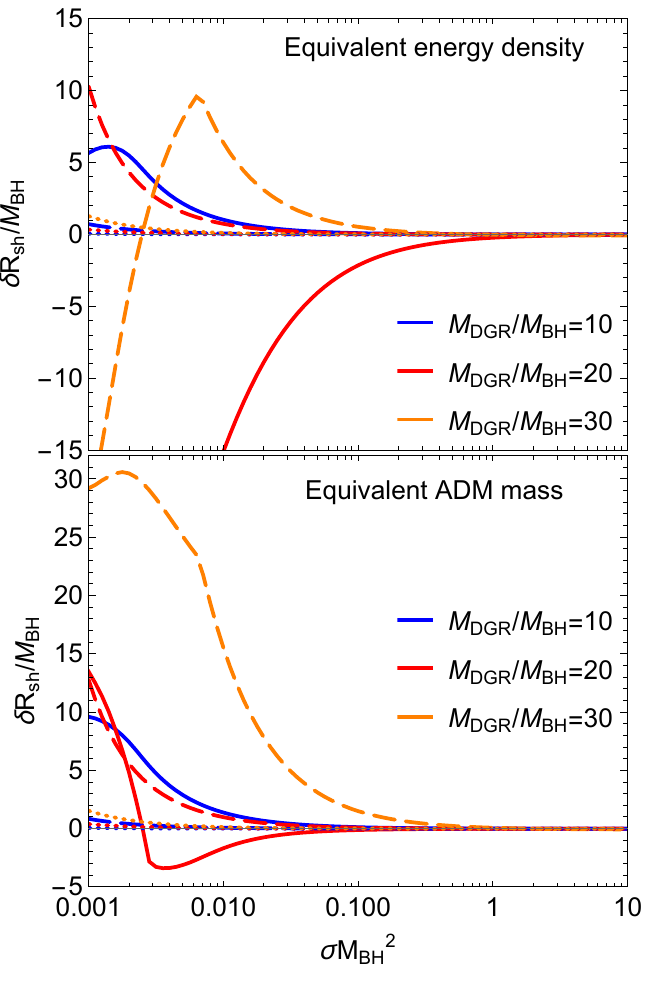}
    \caption{Discrepancy in the shadow radius between classical GR and finite brane tension for (top) equivalent energy density distributions and (bottom) equivalent ADM mass. For the classical GR case ($\sigma\to\infty$), $\delta R_{sh}=0$. The solid, dashed, and dotted lines represent configurations with $a_0/M_{BH}=10$, $a_0/M_{BH}=20$, and $a_0/M_{BH}=40$, respectively. In the most compact DM configurations, the presence of a DM-induced photon sphere leads to behavior distinct from less compact cases.}
    \label{fig. delta Rsh}
\end{figure}

Since our setup is not directly related to current BH shadow observations (e.g. Sagittarius A* or M87*), we instead examine the discrepancy between the shadow radius in the braneworld and classical GR scenarios,
\begin{equation}
    \delta R_{sh}=R_{sh}^{(BW)}-R_{sh}^{(GR)},
\end{equation}
where $R_{sh}^{(BW)}$ and $R_{sh}^{(GR)}$ denote the shadow radius in the braneworld and classical GR cases, respectively. A positive (negative) discrepancy corresponds to a larger (smaller) shadow radius in the braneworld scenario. We consider two types of comparisons:

\begin{enumerate}
    \item \textbf{Equivalent energy density distribution}: The shadow radius in the classical GR case is computed using an equivalent $\epsilon_0$ as in the braneworld case, related to $M_{DGR}$ by Eq.~\eqref{eq. MDGR}. In this comparison, since the finite brane tension introduces an ADM mass discrepancy, the ADM masses of the systems in the finite brane tension and classical GR cases are generally unequal.
    
    \item \textbf{Equivalent ADM mass}: The shadow radius in the classical GR case is computed using an equivalent ADM mass, particularly for the DM mass contribution. In other words, for the discrepancy between the braneworld and classical GR scenarios with $\sigma=\sigma_e$, the shadow radius in the classical GR scenario is computed with
    \begin{equation}
    \epsilon_0=\frac{4\tilde{M}_{DGR}(a_0+r_c)}{\kappa^2a_0^4},
    \end{equation}
    where $\tilde{M}_{DGR}\equiv\lim_{r\to\infty}\left.m(r)\right|_{\sigma=\sigma_e}-M_{BH}$, while we use $\epsilon_0$ given by Eq.~\eqref{eq. MDGR} for the finite brane tension case. This ensures that the systems in both the braneworld and classical GR cases have equivalent total ADM mass. The motivation for this is that the shadow radius is a direct observable. While one may be unable to measure the matter content of such a system, the total mass can be inferred—hence, it is important to compare observables based on measurable quantities.
\end{enumerate}
We show the discrepancy as a function of the brane tension in Fig.~\ref{fig. delta Rsh}. One may refer to Tab.~\ref{tab. Rsh ER GR} as a reference for the BH shadow radius in the classical GR case for each configuration. 

Two distinct behaviors of the discrepancy are observed. In most cases, smaller brane tension tends to produce a positive discrepancy. This effect is caused by the reduced peak of the effective potential at the BH photon sphere as the brane tension decreases, given that the BH photon sphere typically has a higher peak than the DM one (if the latter exists). On the other hand, more compact DM configurations exhibit a decreasing shadow radius, leading to a negative discrepancy as the brane tension decreases. This is due to the DM photon sphere having a higher peak than that of the BH, with the peak increasing as the brane tension increases. As a consequence, we observe a change in trend for configurations where both photon spheres coexist (see the yellow-dashed curve in Fig.~\ref{fig. delta Rsh}).

This is an interesting phenomenon arising from the braneworld theory. Previously, we found that smaller brane tension reduces both the ADM mass and the overall Misner–Sharp mass function. Intuitively, one might expect the BH shadow radius to decrease with smaller brane tension—but instead, it increases. It is also shown that, under equivalent ADM mass conditions, the overall shadow radius becomes larger as the brane tension decreases. Combining these results with supplementary observables would be valuable in determining whether the brane tension can be constrained through optical observations.

\subsection{Einstein ring radius}
\label{sec. einstein ring radius}
The presence of DM around the BH leads to stronger light deflection at large distances compared to an isolated BH. One of the lensing observables is the \emph{Einstein ring}~\cite{Wambsganss:1998gg}, a circular image of a light source seen by the observer as a result of light bending near a massive object. The condition for Einstein ring formation is that the source, the lens, and the observer are collinear, or $\alpha=0$ (c.f. Fig.~S.1 in the supplementary material of Ref.~\cite{Fauzi:2025yse}). As a consequence, the angles of the primary and secondary images are equal, i.e. $\Theta^{(+)}=\left|\Theta^{(-)}\right|=\Theta$. With standard geometry, the impact parameter is related to the image angle by
\begin{equation}
    \tan\Theta=\frac{b}{D_{OL}}. \label{eq. angular rad b}
\end{equation}
As this observable occurs in a weaker-field regime, we use it as a supplement to the shadow radius results obtained previously.

\begin{figure}[htbp!]
    \centering
    \includegraphics[width=0.8\textwidth]{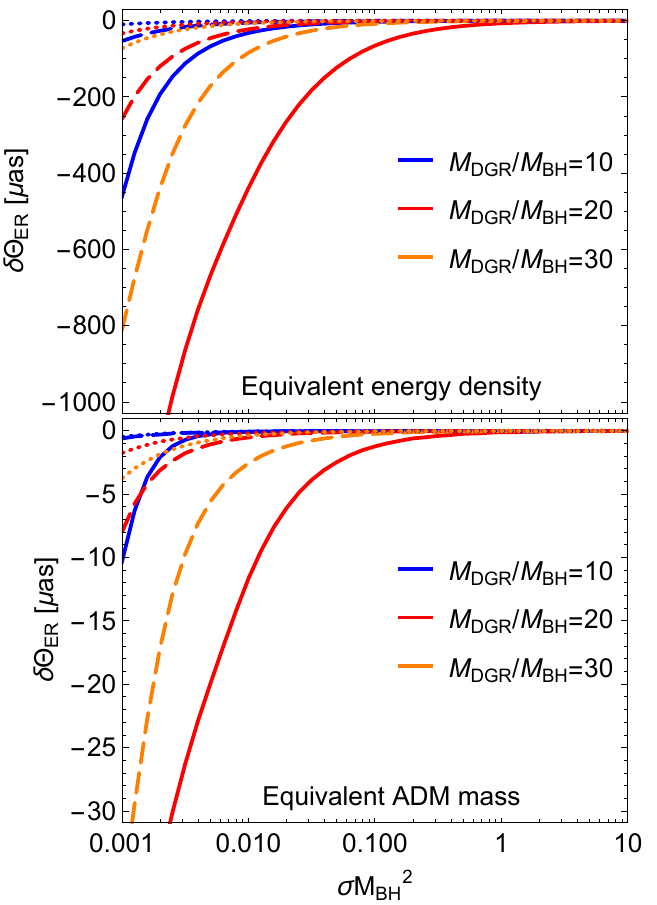}
    \caption{Discrepancy of Einstein ring radius between GR and finite brane tension with (top) equivalent energy density distribution and (bottom) equivalent ADM mass; for the classical GR case ($\sigma\to\infty$), $\delta \Theta_{ER}=0$. The solid, dashed, and dotted lines represent configurations with $a_0/M_{BH}=10$, $a_0/M_{BH}=20$, and $a_0/M_{BH}=40$, respectively. In the equivalent energy density comparison, all configurations follow a trend similar to the ADM mass discrepancy shown in Fig.~\ref{fig. delta adm}. However, a slight discrepancy of lower order of magnitude is observed in the equivalent ADM mass comparison.}
    \label{fig. delta ER}
\end{figure}

We employ a full numerical integration of the geodesic equation to compute the lensing quantities. Following a ray-tracing procedure, we integrate the differential equation given by Eq.~\eqref{eq. radial geodesic equation}. The integration begins from the observer at $D_{OL}$ toward the BH, up to the point of closest approach $r_0$. From there, the integration continues to the source located at $D_{LS}$. This procedure yields the quantity $\phi_{end}$, given by
\begin{equation}
    \phi_{end}(b)=-\int_{D_{OL}}^{r_0}\left(\frac{d\phi}{dr}\right)dr+\int_{r_0}^{D_{LS}}\left(\frac{d\phi}{dr}\right)dr,
    \label{eq. phi end}
\end{equation}
where $d\phi/dr$ is given in Eq.~\eqref{eq. radial geodesic equation}. The minus sign in the first integral accounts for the backward integration. Since the Einstein ring occurs when $\alpha=0$, one requires
\begin{equation}
    \phi_{end}(b_{ER})=\pi, \label{eq. phi end ER}
\end{equation}
where $b_{ER}$ is the photon impact parameter that produces an Einstein ring. We then solve for $b_{ER}$ using Eq.~\eqref{eq. phi end ER} with a standard bisection root-finding method, and the angular radius of the Einstein ring $\Theta_{ER}$ follows from Eq.~\eqref{eq. angular rad b}.

One must specify the choices of $D_{OL}$ and $D_{LS}$ to compute the integral in Eq.~\eqref{eq. phi end}. We follow the angular resolution capabilities of current telescopes as a reference, specifically the EHT, with an angular resolution of $\sim 19\,\mu\text{as}$~\cite{EventHorizonTelescope:2024xos}. For the shadow radius of a Schwarzschild BH ($3\sqrt{3}M_{BH}$), this limit is achieved when\footnote{As a reference, a $19\,\mu\text{as}$ shadow angular radius of a Schwarzschild BH with mass $\sim 1\,M_{\odot}$ corresponds to a BH located at a distance of $\sim 0.003\,\text{pc}$ from Earth.} $D_{OL}\approx5.6\times10^{10}M_{BH}$. We therefore choose $D_{OL}=10^{10}M_{BH}$. For the source distance, we set $D_{LS}=10^{3}M_{BH}$ so that the DM effect remains significant in the lensing process. If the source distance is too large, the geometry along the line of sight would effectively reduce to Schwarzschild spacetime.

Similar to our previous analysis of the BH shadow radius, we compute the discrepancy of the Einstein ring angular radius between the finite brane tension and classical GR cases. The quantity is given by
\begin{equation}
    \delta \Theta_{ER}=\Theta_{ER}^{(BW)}-\Theta_{ER}^{(GR)},
\end{equation}
where $\Theta_{ER}^{(BW)}$ and $\Theta_{ER}^{(GR)}$ are the Einstein ring angular radii in the braneworld and classical GR scenarios, respectively. One may also refer to Tab.~\ref{tab. Rsh ER GR} for the Einstein ring angular radius in classical GR.

The Einstein ring radius behaves as expected intuitively. Within the equivalent DM energy density setup, its value decreases with decreasing brane tension, following a trend similar to the decrease in the ADM mass (c.f. Fig.~\ref{fig. delta adm}). This suggests that the total ADM mass of the system may still be inferred from lensing observables to some degree of precision. Meanwhile, results from the equivalent ADM mass setup reveal noticeable discrepancies in the Einstein ring radius, albeit at significantly smaller angular scales. Given that this observable exhibits a different trend compared to the BH shadow radius, a combination of both observables may provide a useful method to constrain the brane tension—\emph{if} such a special astrophysical scenario can be realized in nature.

\section{Conclusion}
\label{sec. conclusion}
In this work, we investigated solutions of a BH surrounded by an astrophysical environment in the braneworld theory. The formalism developed by Nakas and Kanti~\cite{Nakas:2020sey} is used to construct the BH solution, which can be readily transformed into a regular BH~\cite{Neves:2021dqx}. Using the effective field equations on the brane, we employ the Einstein cluster description for the environment surrounding the BH, with a DM energy density profile~\cite{Cardoso:2021wlq}. The braneworld and regular BH modification is characterized by the brane tension $\sigma$ and the regularization parameter $\ell$ with a dimension of length, respectively. The classical GR limit is recovered when $\sigma\to\infty$ and $\ell\to0$.

We found several intriguing features regarding the properties of the Einstein cluster arising from the braneworld modification. The effective energy density sourcing the four-dimensional spacetime geometry is reduced when non-zero anisotropy is present, with a more significant reduction in the high-anisotropy regime. Since the Einstein cluster has vanishing radial pressure, the reduction in the effective energy density is guaranteed, making the ADM mass of the Einstein cluster in the braneworld smaller than that in classical GR for the same density distribution. Moreover, a recursive relation between the transverse pressure and matter compactness leads to an increase in anisotropy at high compactness, which prevents horizon formation in the Einstein cluster. As a consequence, one may, in principle, obtain an arbitrarily dense anisotropic matter distribution without forming a BH. Similar behavior was also observed in Ref.~\cite{Lugones:2017jak} for neutron stars, where no maximum mass exists in the presence of finite brane tension. We also find that the energy conditions and the speed of sound are modified as the brane tension varies, with the DEC and causal limit violated for ultracompact DM configurations.

The effective energy density in Eq.~\eqref{eq. epsilon eff} provides interesting insights into the effects of finite brane tension on anisotropic objects. In the case of an anisotropic gravastar with an ansatz for anisotropy as given in Refs.~\cite{DeBenedictis:2005vp,Fauzi:2024nta}, we expect similar behavior in both the spacetime geometry\textemdash as well as the matter properties\textemdash as observed in our Einstein cluster results: horizon formation by anisotropic stars may also be prevented if the anisotropy increases proportionally with the local compactness. Although this scenario is interesting to explore, solving the field equations would require significantly more effort, since the radial pressure is now non-zero, leading to additional contributions in Eqs.~\eqref{eq. epsilon eff}, \eqref{eq. p eff}, \eqref{eq. pt eff}, etc.

We further examined the optical observables arising from this astrophysical scenario, namely the BH shadow radius and the Einstein ring angular radius. Since the considered setup is not directly relevant to current observations, we focus on the differences between the finite brane tension case and the classical GR case. Our results reveal that, despite the decrease in ADM mass, the overall BH shadow radius increases as the brane tension decreases. In contrast, the Einstein ring angular radius exhibits a different behavior, varying almost linearly with the decreasing ADM mass. Since these two observables show distinct trends with respect to the brane tension, one may, at least in principle, constrain the brane tension by combining measurements of both quantities. Of course, achieving such measurements in the near future is unlikely, as no observational evidence for such a scenario currently exists. Nevertheless, our results provide useful insights into possible optical signatures of finite brane tension.

We also highlight several directions for future work. Such modifications are expected to produce distinct signatures in gravitational wave signals; however, since gravitational waves may propagate through higher dimensions (i.e. the bulk), one must incorporate the effects of the extra dimension. For example, to compute the quasinormal modes of the BH, one should use the master equation introduced in~\cite{Kodama:2003jz}, rather than the standard Regge–Wheeler–Zerilli equation. Despite the increased complexity of the master wave equation, gravitational perturbations may not propagate deeply into the bulk~\cite{Abdalla:2001he,Toshmatov:2016bsb}, suggesting that bulk effects could be subdominant. Indeed, gravitational perturbations of tidally charged braneworld BHs have been studied in Ref.~\cite{Toshmatov:2016bsb}, where the effective potential for axial perturbations takes a form similar to that of the Regge–Wheeler equation. Additionally, one may investigate curvature scalars arising from the modified DM spacetime geometry to determine whether curvature singularities still emerge, even in the absence of horizon formation.




\paragraph{Data/Software/Code Availability Statements:} This manuscript is purely theoretical. No new data were generated or analyzed during this study. The code used for the derivations and calculations is available from the corresponding author upon reasonable request.


\bibliographystyle{JHEP}
\bibliography{biblio.bib}


\end{document}